\documentclass[12pt]{JHEP3}
\usepackage{amsmath}
\usepackage{graphicx}
\usepackage{multirow}
\usepackage{epsfig}
\usepackage{rotating}

\catcode`@=11
\def\@citex[#1]#2{\if@filesw\immediate\write\@auxout{\string\citation{#2}}\fi
  \def\@citea{}\@cite{\@for\@citeb:=#2\do
    {\@citea\def\@citea{,\penalty\@m}\@ifundefined
      {b@\@citeb}{{\bf ?}\@warning
       {Citation `\@citeb' on page \thepage \space undefined}}%
\hbox{\csname b@\@citeb\endcsname}}}{#1}}

\def\citer{\@ifnextchar
[{\@tempswatrue\@citexr}{\@tempswafalse\@citexr[]}}
\def\@citexr[#1]#2{\if@filesw\immediate\write\@auxout{\string\citation{#2}}\fi
  \def\@citea{}\@cite{\@for\@citeb:=#2\do
    {\@citea\def\@citea{--\penalty\@m}\@ifundefined
       {b@\@citeb}{{\bf ?}\@warning
       {Citation `\@citeb' on page \thepage \space undefined}}%
\hbox{\csname b@\@citeb\endcsname}}}{#1}}
\catcode`@=12

\title{\bf Improved Constraints on $Z'$ Bosons from Electroweak Precision Data}
\author{Jens Erler$^1$\footnote{erler@fisica.unam.mx} , 
Paul Langacker$^2$\footnote{pgl@ias.edu} , 
Shoaib Munir$^1$\footnote{smunir@fisica.unam.mx}  and
Eduardo Rojas$^1$\footnote{eduardor@fisica.unam.mx}
\\ \\
{\normalsize\it $^1$Departamento de F\'isica Te\'orica, Instituto de F\'isica,} \\ 
{\normalsize\it Universidad Aut\'onoma de M\'exico, 04510    M\'exico D.F., M\'exico} \\
{\normalsize\it $^2$School of Natural Sciences, Institute for Advanced Study,} \\ 
{\normalsize\it Einstein Drive, Princeton, NJ 08540, USA}}
\newcommand{\be}{\begin{equation}}
\newcommand{\ee}{\end{equation}}
\newcommand{\ba}{\begin{array}}
\newcommand{\ea}{\end{array}}
 
\date{}

\abstract{We analyze various models with an extra $U(1)$ gauge symmetry in addition to the Standard Model (SM) gauge group at low energies, and impose limits on the mass of the neutral $Z'$ boson, $M_{Z'}$, predicted in all such models, and on the $Z-Z'$ mixing angle, $\theta_{ZZ'}$. The precision electroweak (EW) data strongly constrain $\theta_{ZZ'}$ to very small values and for most models we find lower limits on $M_{Z'}$ of ${\cal O}(1~{\rm TeV})$.  In one case we obtain a somewhat better fit than in the SM (although this is only marginally statistically significant) and here we find a weak upper limit at the 90\% C.L.}

\begin{document}



\section{Overview}
\label{intro}
With the advent of the LHC, particle physics has entered a new exciting era. Within a few years of data accumulation, the LHC should be able to test and constrain many types of new physics beyond the SM. In particular, the discovery reach for extra neutral gauge bosons is exceptional.  Searches for a high invariant dilepton mass peak in about 100 fb$^{-1}$ of accumulated data will find or exclude $Z'$ bosons up to about 5~TeV, and a luminosity upgraded LHC (by roughly a factor of 10) can extend the reach by another TeV~\cite{Godfrey:2002tna}. After a potential discovery, the LHC will have some diagnostic means to narrow down the underlying $Z'$ model~\cite{Petriello:2008zr} by studying, for example, leptonic forward-backward asymmetries (for reviews, see Refs.~\citer{Leike:1998wr,Langacker:2008yv}) and heavy quark final states~\cite{Barger:2006hm,Godfrey:2008vf}. Furthermore, angular distributions of Drell-Yan leptons may help to discriminate a $Z'$ against spin-0 (sneutrino) and spin-2 (Kaluza-Klein graviton) resonances~\cite{Osland:2009tn}. However, the hadronic LHC environment will make it difficult to specify the $Z'$ properties completely or with satisfactory precision. Electroweak precision measurements will therefore play an important complementary r$\hat{\rm o}$le in this context. Already these data give rise to strong constraints on possible $Z'$ and in many cases yield the best limits.  In this paper we will revisit the EW precision data in the presence of $Z'$ bosons. This is motivated by much higher precision in several measurements (such as from the Tevatron) compared to previous studies~\cite{Erler:1999ub,Erler:1999nx} and also by significant shifts and new measurements at low energies, a sector with increasing impact on global analyses of this type.

Neutral gauge sectors with an additional $U(1)'$ symmetry in addition to the SM hypercharge $U(1)_Y$ and an associated $Z'$ gauge boson are among the best motivated extensions of the SM. They are predicted in most Grand Unified Theories (GUTs) and appear copiously in superstring theories. An especially compelling motivation for extended gauge theories came from the development of GUTs larger than the original $SU(5)$ model, such as $SO(10)$ or $E_6$, which allow the SM gauge group to be embedded into them~\cite{Georgi:1974sy}. There is an extensive range of models with an extra $U(1)$ symmetry (for a review, see~\cite{Langacker:2008yv} and references therein). Among these, models based on the $E_6$ GUT group and left-right symmetry groups have been extensively pursued in the literature and are particularly significant from the point of view of LHC phenomenology. In the context of supersymmetry, this class of models also arises~\cite{Erler:2000wu} from requiring the extra $U(1)'$ gauge symmetry to provide a solution to the $\mu$-problem~\cite{Kim:1983dt}, to forbid rapid (dimension~4) proton decay, to protect all fields by chirality and supersymmetry from acquiring high scale masses, and to be consistent with anomaly cancellation, gauge coupling unification and family universality (to avoid the strong constraints from the flavor-changing neutral current (FCNC) sector~\cite{Langacker:2000ju}). The models studied here include: 

\begin{description}
\item[$E_6$ based models:]
Extra $U(1)$ gauge symmetries appear in the decomposition of the $SO(10)$ \cite{Robinett:1982tq} or $E_6$~\cite{Langacker:1984dc,Hewett:1988xc} GUT groups. $E_6$ contains the maximal subgroup $SO(10) \times U(1)_\psi$, and $SO(10)$ can be further decomposed into its $SU(5) \times U(1)_\chi$ maximal subgroup. We are considering models in which the linear combination,
\begin{eqnarray}
U(1)' = \cos\beta\ U(1)_\chi  + \sin\beta\ U(1)_\psi,
\label{E6}
\end{eqnarray} 
survives down to the EW scale, using a convention in which the mixing angle in Eq.~(\ref{E6}) satisfies\footnote{Note, that by restricting $\beta$ to a semi-circle makes the sign of the mixing angle $\theta_{ZZ''}$ physically meaningful.} $-90^\circ < \beta \leq 90^\circ$. The full $E_6$ symmetry would impose strong constraints on these models, which are often unrealistic.  For the purpose of our analyses, we are, however, mostly interested in the effects, phenomenology, and constraints associated directly with the $Z'$ bosons and not in other aspects of these models\footnote{There are classes of $Z'$ models closely related to the $E_6$ ones motivated by minimal gauge unification~\cite{Langacker:2008yv,Erler:2000wu} which generically require more than one kind of $U(1)'$ breaking singlet field. The issue of obtaining a suitable symmetry breaking pattern is discussed in~\cite{Langacker:2008dq}.}.

\begin{itemize}
\item [$Z_\chi$:]
$\beta = 0^\circ \Longrightarrow Z' = Z_\chi$, which is also the unique solution to the conditions of (i) family universality, (ii) no extra matter other than right handed neutrino, (iii) absence of gauge and mixed gauge/gravitational anomalies and (iv) orthogonality to the hypercharge generator.  

\item[$Z_\psi$:]
$\beta = 90^\circ \Longrightarrow Z' = Z_\psi$, possessing only axial vector couplings to ordinary fermions. As discussed in Section~\ref{results}, it is among the least constrained by the precision data.

\item[$Z_\eta$:]
$\beta = - \arctan \sqrt{5/3} \approx -52.2^\circ \Longrightarrow Z' =  \sqrt{3/8}\, Z_\chi - \sqrt{5/8}\, Z_\psi \equiv Z_\eta$, occuring in Calabi-Yau compactifications~\cite{Candelas:1985en} of the heterotic string~\cite{Gross:1984dd} if $E_6$ breaks directly to a rank 5 subgroup~\cite{Witten:1985xc} via the Hosotani mechanism~\cite{Hosotani:1983vn}.

\item[$Z_I$:]
$\beta = \arctan \sqrt{3/5} \approx 37.8^\circ  \Longrightarrow Z' = \sqrt{5/8}\, Z_\chi + \sqrt{3/8}\, Z_\psi \equiv - Z_I$, which is orthogonal to the $Z_\eta$. This boson~\cite{Langacker:1984dc} has the defining property of vanishing couplings to up-type quarks.  Its production is thus suppressed at hadron colliders, especially at the Tevatron since in high-energy $p\bar{p}$ collisions $Z'$ production through down quarks is suppressed by a factor of 25 relative to up quarks~\cite{Leike:1998wr}.

\item[$Z_S$:] 
A supersymmetric model with a secluded $U(1)'$ breaking sector and a large supersymmetry breaking $A$-term was introduced (i) to provide an approximately flat potential allowing the generation of a $Z$--$Z'$ mass hierarchy~\cite{Erler:2002pr} and (ii) to produce a strong first order EW phase transition for EW baryogenesis~\cite{Kang:2004pp}. Such a sector is obtained if the right-handed $\nu$ has $-2\times$ the $U(1)'$ charge of the other SM singlet in a ${\bf 27}$ representation of $E_6$, giving $\beta = \arctan \sqrt{5/27} \approx 23.3^\circ \Longrightarrow Z' = \sqrt{27/32}\, Z_\chi + \sqrt{5/32}\, Z_\psi \equiv Z_S$, which is numerically close to the $Z_I$. 
  
\item[$Z_N$:]
$\beta = \arctan \sqrt{15} \approx 75.5^\circ \Longrightarrow Z' = (Z_\chi + \sqrt{15}\, Z_\psi)/4 \equiv Z_N$, which is a consequence of
choosing the right-handed neutrinos to have zero $U(1)'$ charges so that they can acquire very heavy Majorana masses~\citer{Ma:1995xk,Kang:2004ix} and are thus suitable to take part in the standard seesaw mechanism~\cite{GellMann:1980vs} with three naturally light neutrinos. The $Z_N$ boson also appears in a model referred to as the ESSM~\cite{King:2005jy} or the E$_6$SSM~\cite{King:2006rh}.  

\item[$Z_R$:]
All models discussed so far assume negligible kinetic mixing, {\em i.e.,\/} the absence of a cross term,
\be
- {\sin \chi\over 2} F_{\mu\nu}' F^{\mu\nu}_Y,
\label{kineticmixing}
\ee
between the gauge kinetic terms for the $U(1)'$ and the $U(1)_Y$ gauge bosons~\citer{Holdom:1985ag,Foot:1991kb}. This is motivated by the orthogonality of all $U(1)$ subgroups within a simple GUT group like $E_6$.  A usually very small kinetic term arises at the two-loop level from the renormalization group evolution of the gauge couplings~\cite{Langacker:1998tc}. An exception in the case in which incomplete GUT multiplets survive below the unification scale and in which a relatively large kinetic mixing term can be generated~\cite{Babu:1996vt}. In any case, there is no general reason to ignore kinetic mixing and we now address it in a different but equivalent formalism: one can always redefine the gauge boson fields to remove any term of the form (\ref{kineticmixing}). The effects due to $\sin\chi \neq 0$ then manifest themselves in the $U(1)'$ charges which will in general have a non-trivial hypercharge component. In the $E_6$ context, for example, one can write the $Z'$ as the general (family-universal) combination~\cite{Erler:1999nx},
\be
Z' = \cos\alpha \cos\beta Z_\chi + \sin\alpha \cos\beta Z_Y + \sin\beta Z_\psi.
\label{generalE6}
\ee
The restriction to $\beta = 0^\circ$ corresponds to general $SO(10)$ based models. Specifying further to $\alpha = \arctan \sqrt{3/2} \approx 50.8^\circ \Longrightarrow Z' = \sqrt{2/5}\, Z_\chi + \sqrt{3/5}\, Z_Y\equiv Z_R$, which couples to charges proportional to the diagonal (third) component of right-handed isospin, $SU(2)_R$. We are unaware of this case having been introduced in the literature.  But as we will discuss in Section~\ref{results}, the resulting $Z'$ gives a reasonably good fit, and technically even a finite 90¼\% C.L. upper limit can be set on its mass.

\item[$Z_{LR}$:]
Models with left-right symmetry (reviewed in Ref.~\cite{Mohapatra:1986uf}) are based on the gauge group $SU(3)_C \times SU(2)_L \times SU(2)_R \times U(1)_{B-L} \subset SO(10)$ and contain a boson, $Z_{LR}  \equiv \sqrt{3/5}\, (\bar\alpha\, Z_R - Z_{B-L}/2\bar\alpha)$. Here $B$ and $L$ coincide, respectively, with baryon and lepton number for the ordinary fermions, and $Z' = \sqrt{3/5}\, Z_\chi - \sqrt{2/5}\, Z_Y \equiv - \sqrt{3/8}\, Z_{B-L} $ is obtained by choosing $\alpha = - \arctan \sqrt{2/3} \approx - 39.2^\circ$ ($B$ or $L$ individually cannot be parametrized in this way). The parameter $\bar\alpha = \sqrt{g_R^2/g_L^2 \cot^2 \theta_W - 1}$, with $\theta_W$ the weak mixing angle, gives the coupling strength in terms of the $SU(2)_{L,R}$ gauge couplings, $g_{L,R}$,. Manifest left-right symmetry (which we will assume) requires $g_L = g_R$, while the very strong coupling limit ($\bar\alpha, g_R/g_L \to \infty$) implies $Z_{LR} \rightarrow Z_R$.

\item[$Z_{\not L}$:] 
A leptophobic $Z'$ has vanishing $U(1)'$ charges to charged leptons and left-handed neutrinos.  One version of this idea~\cite{Babu:1996vt} is a variation of the $Z_\eta$ model with kinetic mixing added. The choice $(\alpha, \beta) = (\arctan \sqrt{8/27}, - \arctan \sqrt{9/7}) \approx (28.6^\circ, - 48.6^\circ ) \Longrightarrow Z' = \sqrt{27/80}\, Z_\chi+ 1/\sqrt{10}\, Z_Y - 3/4\, Z_\psi \equiv Z_{\not L}$. The effects of a leptophobic $Z'$ are very difficult to observe but it can be searched for in the dijet~\cite{Aaltonen:2008dn} and $t\bar{t}$ \cite{Aaltonen:2007dz} channels at hadron colliders. Moreover, mixing effects at LEP~1 strongly constrain $\theta_{ZZ'}$ even in this case.  
\end{itemize}

\item[Sequential $Z'$:]
The $Z_{SM}$ boson is defined to have the same couplings to fermions as the SM $Z$. Such a boson is not expected in the context of gauge theories unless it has different couplings to exotic fermions than the ordinary $Z$. However, it serves as a useful reference case when comparing constraints from various sources. It could also play the role of an excited state of the ordinary $Z$ in models of compositeness or with extra dimensions at the weak scale.

\item[A superstring $Z'$:]
There is a family non-universal $Z_{string}$ boson appearing in a specific model~\cite{Chaudhuri:1994cd} based on the free fermionic string construction with real fermions. This model has been investigated in considerable detail~\cite{Cleaver:1997jb,Cleaver:1998gc} with the goal of understanding some of the characteristics of (weakly coupled) string theories, and of contrasting them with the more conventional ideas such as GUTs. While this specific model itself is not realistic (for example, it fails to produce an acceptable fermion mass spectrum) the predicted $Z_{string}$ it contains is itself not ruled out (ignoring issues related to CP violation and FCNCs~\cite{Langacker:2000ju}). Its coupling strength is predicted and so are its fermion couplings. Such a $Z_{string}$ can be naturally at the electroweak scale~\cite{Cvetic:1995rj,Cvetic:1997ky}.
\end{description}


\section{Extended Higgs Sectors and Exotics}
\label{exotics}
The incorporation of (one or more) extra gauge group(s) in the models listed above generally warrants an extended fermionic sector for two main reasons: (i) cancellation of gauge and mixed gauge-gravitational anomalies to assure quantum consistency of the theory, and (ii) in the context of low-energy supersymmetry, the unification of gauge couplings at high energies. Provided that all fermions belong to complete $E_6$ representations the anomalies are cancelled automatically. In a bottom-up approach, however, the condition of anomaly cancellation restricts the $U(1)'$ charge assignments of the SM fermions and the exotics~\cite{Erler:2000wu,Appelquist:2002mw}. 

The structure of the Higgs sector of the underlying model is important as it may affect the $\rho_0$ parameter,
\be
\rho_0 \equiv \frac{\sum_i (t_i^2 - t_{3i}^2 + t_i) |\langle \phi_i \rangle|^2}
                {\sum_i 2 t_{3i}^2               |\langle \phi_i \rangle|^2},
\label{rho}
\ee
where $t_i$ ($t_{3i}$) is (the third component of) the weak isospin of the Higgs field $\phi_i$, and which enters the neutral and charged ($M_W$) gauge boson mass interdependence,
\begin{eqnarray}
   M_0 = {M_W\over \sqrt{\rho_0} \cos\theta_W}.
\label{wzrelation}
\end{eqnarray}
$\rho_0 = 1$  corresponds to a Higgs sector with only $SU(2)$ doublets and singlets. In that case the mass parameter $M_0$ (the ordinary $Z$ mass in the absence of $Z-Z'$ mixing) is predicted. In general there is mixing between the mass eigenstates of the $Z'$ and the $Z$ given by~\cite{Langacker:1984dp},
\begin{eqnarray}
\tan^2 \theta_{ZZ'} =\frac{M_0^2-M_Z^2}{M_{Z'}^2-M_0^2}.
\end{eqnarray}

Allowing $\rho_0$ as an additional fit parameter means that the Higgs sector of the model is arbitrary and may include higher-dimensional Higgs representations. In addition, the presence of non-degenerate multiplets of heavy fermions or scalars will affect the $W$ and $Z$ self energies at the loop level, and therefore contribute to the $T$ parameter~\cite{Peskin:1990zt}. With the current data set, the phenomenological consequences of $\rho_0$ and $T$ are indistinguishable and values quoted for $\rho_0$ really apply to the combination $\rho_0/(1-\alpha T)$. 

\begin{table}[t!]
\begin{tabular}{cccc}  
\hline  & &\vspace{-8pt} \\
$Z'$ & $C$ & range &restricted range $\left( \omega = 0, \tau \geq  -  \hspace{-12pt}  \ba{c} 1 \\ 2 \ea \hspace{-4pt} \right)$ \vspace{8pt} \\
\hline
 & &\vspace{-8pt} \\
$Z_\chi$ &$ \ba{c} 2 \\ \overline{\sqrt{10}} \ea \left( 1 - - \hspace{-12pt}  \ba{c} 5 \\ 2 \ea \omega \right) $ &
$ \ba{c} 2 \\ \overline{\sqrt{10}} \ea \left[ - - , \hspace{-18pt}  \ba{c} 3 \\ 2 \ea +1 \right]$ &
$ \ba{c} 2 \\ \overline{\sqrt{10}} \ea$ \vspace{8pt} \\
$Z_\psi$ &$ \sqrt{ \ba{c} 2 \\ \overline{ \hspace{2pt} 3 \hspace{2pt} } \ea}  \left(1 - 2\ \tau - -  \hspace{-12pt}  \ba{c} 3 \\ 2 \ea \omega \right)$ &  
$ \sqrt{  -  \hspace{-12pt} \ba{c} 2 \\ 3 \ea}  \left[ -1, +1 \right]$ &
$ \sqrt{  -  \hspace{-12pt} \ba{c} 2 \\ 3 \ea}  \left[ -1, 0  \right]$ \vspace{8pt} \\
$Z_\eta$ &$ - \ba{c} 1 \\ \overline{\sqrt{15}} \ea \left( 1 - 5\ \tau \right) $ & 
$ \ba{c} 1 \\ \overline{\sqrt{15}} \ea  \left[ -1, +4 \right]$ & 
$ \ba{c} 1 \\ \overline{\sqrt{15}} \ea  \left[  - , \hspace{-18pt}  \ba{c} 3 \\ 2 \ea + 4 \right]$ \vspace{8pt} \\
$Z_I$ & $\tau + 2\ \omega -1$ & $[-1,1]$ & $\left[- - \hspace{-15pt}  \ba{c} 1 \\ 2 \ea, 0 \right]$ \vspace{8pt} \\
$Z_S$  & $ \ba{c} 7 \\ \overline{\sqrt{60}} \ea \left( 1 - -  \hspace{-12pt}  \ba{c} 5 \\ 7 \ea \tau -  -\hspace{-15pt}  \ba{c} 15 \\ 7 \ea\omega \right) $ & $ \ba{c} 7 \\ \overline{\sqrt{60}} \ea  \left[- - , \hspace{-18pt}  \ba{c} 8 \\ 7 \ea +1 \right]$ &
$ \ba{c} 1 \\ \overline{\sqrt{15}} \ea  \left[+ 1, + - \hspace{-15pt}  \ba{c} 9 \\ 4 \ea  \right]$ \vspace{8pt} \\
$Z_N$    & $ \ba{c} 3 \\ \overline{\sqrt{10}} \ea \left( 1 - -  \hspace{-12pt}  \ba{c} 5 \\ 3 \ea (\tau + \omega) \right) $ & 
$ \ba{c} 3 \\ \overline{\sqrt{10}} \ea  \left[- - , \hspace{-18pt}  \ba{c} 2 \\ 3 \ea +1 \right]$ &
$ \ba{c} 1 \\ \overline{\sqrt{10}} \ea  \left[- 2, + - \hspace{-15pt}  \ba{c} 1 \\ 2 \ea  \right]$ \vspace{8pt} \\
$Z_R$    &  $1 - \omega$   &$ [0,+1]$ & $1$  \vspace{8pt} \\
$Z_{LR}$ & \hspace*{3pt}$\sqrt{ \ba{c} 3 \\ \overline{  \hspace{2pt} 5 \hspace{2pt}} \ea} \bar\alpha \left[ 1 - \left(1 +  \hspace*{-4pt} \ba{c} 1 \\ \overline{\bar\alpha^2} \ea \hspace*{-4pt} \right) \omega \right] $ & 
\hspace*{3pt} $ \sqrt{  -  \hspace{-12pt} \ba{c} 3 \\ 5 \ea}  \left[ - - , \hspace{-18pt} \ba{c} 1 \\ \bar\alpha \ea +\bar\alpha \right]$ &
$\sqrt{ \ba{c} 3 \\ \overline{  \hspace{2pt} 5 \hspace{2pt}} \ea} \bar\alpha$ \vspace{8pt} \\
$Z_{\not L}$ &$ \sqrt{ \ba{c} 3 \\ \overline{ \hspace{2pt} 2 \hspace{2pt} } \ea}  \tau$ & 
$\sqrt{ \ba{c} 3 \\ \overline{ \hspace{2pt} 2 \hspace{2pt} } \ea}  \left[ 0, +1 \right]$ &
$\sqrt{ \ba{c} 3 \\ \overline{ \hspace{2pt} 2 \hspace{2pt} } \ea}  \left[  - \hspace{-12pt}   \ba{c} 1 \\ 2 \ea , +1 \right]$ \vspace{8pt} \\
\hline
\end{tabular}
\caption{Special Higgs sectors for $E_6$ based models. The third column shows the most general range for $C$ if all three Higgs doublets in a {\bf 27} representation participate in spontaneous symmetry breaking. The last column corresponds to the restricted range appropriate for supersymmetry inspired models.}
\label{cranges}
\end{table}

If the $U(1)'$ charge assignments of the Higgs fields, $Q'_i$, are known in a specific model, then there exists an additional constraint~\cite{Langacker:1991pg},
\begin{eqnarray}
  \theta_{ZZ'} = C\ {g_2\over g_1} {M_Z^2\over M_{Z^\prime}^2},
\end{eqnarray}
where $g_1 = g_L/\cos\theta_W$ and where $g_2 = \sqrt{5/3}\ g_1 \sin\theta_W \sqrt{\lambda}$ is the $U(1)^\prime$ gauge coupling. The latter is given in terms of $\lambda$ which is of order unity (we will set $\lambda = 1$ as is conventionally done), and in fact $\lambda \sim 1$ in GUT models breaking directly to $SU(3)_C \times SU(2)_L \times U(1)_Y \times U(1)'$. $C$ is a function of vacuum expectation values (VEVs) of the Higgs fields and the $Q'_i$,
\be
  C = - \frac{\sum_i t_{3i} Q^\prime_i |\langle \phi_i \rangle|^2}{\sum_i t_{3i}^2 |\langle \phi_i \rangle|^2}.
\label{paraC}
\ee
As an illustration, for the $E_6$ based models one may restrict oneself to the case where the Higgs fields arise from a {\bf 27} representation. The $U(1)'$ quantum numbers are then predicted and Eq.~(\ref{paraC}) receives contributions from the VEVs of three Higgs doublets, $x \equiv \langle\phi_\nu \rangle$, $v \equiv \langle \phi_N \rangle$ and $\bar{v} \equiv \langle \phi_{\bar{N}} \rangle$, respectively, in correspondence with the standard lepton doublet, as well as the two doublets contained in the ${\bf \overline{5}}$ and {\bf 5} of $SU(5) \subset E_6$. They satisfy the sum rule, $|v|^2 + |\bar{v}|^2 + |x|^2 = (\sqrt{2}\ G_F)^{-1} = (246.22 \mbox{ GeV})^2$, and we introduce the ratios,
\be
\tau = {|\bar{v}|^2 \over |v|^2 + |\bar{v}|^2 + |x|^2}  \hspace*{50pt } (0 \leq \tau \leq 1),
\ee
\be
\omega = {|x|^2 \over |v|^2 + |\bar{v}|^2 + |x|^2} \hspace*{50pt } (0 \leq  \omega \leq 1).
\ee
In supersymmetric models one usually assumes $x = \omega = 0$ to avoid spontaneous breaking of lepton number and problems with charged current universality, as well as $\bar{v} \geq v$ (implying $\tau \geq 1/2$) to avoid non-perturbative values for the top quark Yukawa coupling. The resulting ranges for $C$ are shown in Table~\ref{cranges}.

\section{Details of the Analyses}
\label{analysis}
The theoretical evaluation uses the special purpose FORTRAN package GAPP~\cite{Erler:1999ug} dedicated to the Global Analysis of Particle Properties. All experimental and theoretical uncertainties are included and their correlations accounted for. All errors have been added in quadrature and in most (but not all) cases been treated as Gaussian. The effects of the $Z'$ bosons are taken into account as first order perturbations to the SM expressions.

\begin{table}[t]
\begin{tabular}{|l|c|c|c|r|}
\hline Quantity & Group(s) & Value & Standard Model & pull \\ 
\hline
$m_t$\hspace{11pt}[GeV] & Tevatron             &$  173.1    \pm 1.4          $&$ 173.1    \pm 1.4      $&$ 0.0$   \\
$M_W$ [GeV]                     & Tevatron              &$  80.432   \pm 0.039  $&$ 80.380  \pm 0.015  $&$ 1.3$   \\
$M_W$ [GeV]                     & LEP 2                   &$  80.376   \pm 0.033  $&                                        &$-0.1$   \\
\hline
$g_L^2 $                             & NuTeV                 &$ 0.3010 \pm 0.0015  $&$  0.3039 \pm 0.0002    $&$-2.0$   \\
$g_R^2 $                            & NuTeV                 &$ 0.0308 \pm 0.0011  $&$  0.0300 $&$ 0.7$   \\
$\kappa $                            & CCFR                  &$   0.5820 \pm 0.0041  $&$  0.5831 \pm 0.0003  $&$-0.3$   \\
$R^\nu$                              & CDHS                  &$   0.3096 \pm 0.0043 $&$   0.3091 \pm 0.0002   $&$ 0.1$   \\
$R^\nu$                              & CHARM               &$   0.3021 \pm 0.0041 $&                                         &$-1.7$   \\
$R^{\bar\nu}$                     & CDHS                  &$   0.384  \pm 0.018    $&$   0.3861 \pm 0.0001  $&$-0.1$   \\
$R^{\bar\nu}$                     & CHARM               &$   0.403  \pm 0.016    $&$                                      $&$ 1.1$   \\
$R^{\bar\nu}$                     & CDHS 1979        &$   0.365  \pm 0.016    $&$   0.3815 \pm 0.0001  $&$-1.0$   \\
\hline
$g_V^{\nu e}$                    &  CHARM II + older  &$  -0.040  \pm 0.015    $&$  -0.0397 \pm 0.0003  $&$ 0.0$   \\
$g_A^{\nu e}$                    &  CHARM II + older  &$  -0.507  \pm 0.014    $&$  -0.5064 \pm 0.0001   $&$ 0.0$   \\
\hline
$Q_W({\rm Tl})$                 & Oxford + Seattle     &$ -116.4    \pm 3.6       $&$-116.8 $&$ 0.1$   \\
$Q_W({\rm Cs})$               &  Boulder                   &$ -73.16   \pm 0.35      $&$ -73.16 \pm 0.03   $&$ 0.0$   \\
$Q_W (e) $                         &  SLAC E158          &$ -0.0403 \pm 0.0053  $&$ -0.0472 \pm 0.0005   $&$ 1.3$   \\
$\cos\gamma\; C_{1d} - \sin\gamma\; C_{1u}$  & Young et al.        &$     0.342 \pm 0.063    $&  $0.3885 \pm    0.0002$  &$-0.7$   \\
$\sin\gamma\; C_{1d} + \cos\gamma\; C_{1u}$ & Young et al.        &$ -0.0285 \pm 0.0043  $&  $-0.0335 \pm  0.0001$  &$ 1.2$   \\
\hline
CKM unitarity                      & KLOE dominated  &$ 1.0000 \pm 0.0006  $& $1$                            & $0.0$    \\
$(g_\mu - 2 -\alpha/\pi)/2$ & BNL E821             &$ 4511.07 \pm 0.74    $&$ 4509.04 \pm 0.09  $&$ 2.7$   \\   
\hline
\end{tabular}
\caption{Non $Z$-pole precision observables from FNAL, CERN, SLAC, JLab, and elsewhere. Shown are the experimental results, the SM predictions, and the pulls.The SM errors are from the parametric uncertainties in the Higgs boson and quark masses and in the strong and electromagnetic coupling constants evaluated at $M_Z$.} 
\label{Non-Zpole}
\end{table}

The most stringent indirect constraints on $M_{Z'}$ come from low-energy weak neutral current experiments displayed in Table~\ref{Non-Zpole} together with other non $Z$-pole observables. The first set shown are the most recent combinations of $M_W$~\cite{EWWG:2008ub} and the top quark mass, $m_t$~\cite{TEVEWWG:2009ec}. 

The second set are effective four-Fermi operator coefficients ($g_{L,R}^2$) and cross section ratios ($\kappa, R^\nu,R^{\bar\nu}$) from neutrino and anti-neutrino deep inelastic scattering ($\nu$-DIS) at FNAL~\cite{Arroyo:1993xx,Zeller:2001hh} and CERN~\cite{Abramowicz:1986vi,Allaby:1986pd}. The NuTeV~\cite{Zeller:2001hh} results are very preliminary.  We have updated Ref.~\cite{Zeller:2001hh} to account for the recently measured strange quark asymmetry~\cite{Mason:2007zz}. The incorporation of other effects like more recent QED radiative corrections~\cite{Diener:2003ss,Arbuzov:2004zr} and parton distribution functions~\cite{Martin:2003sk}  (allowing some level of charge symmetry violation) are likely to decrease the $2\sigma$ deviation in $g_L^2$ shown in the Table. On the other hand, the world average~\cite{Amsler:2008zzb} of the $K_{e3}$ branching fraction has been corrected upwards several times in the previous years, making for a larger correction for the $\nu_e$ ($\bar\nu_e$) contamination of the dominantly $\nu_\mu$ ($\bar\nu_\mu$) beams, and which by itself would be indicative of an increase in the deviation. More precise statements about the size and the sign of the net effect of these corrections will only be possible after the completion of the re-analysis of the NuTeV result, which is currently in progress~\cite{Zeller:2008}. The $g_{V,A}^{\nu e}$ in the third set are effective four-Fermi couplings for elastic $\nu$-$e$ scattering~\cite{Vilain:1994qy}. 

$Q_W$ denote so-called weak charges measured in atomic parity violation~\citer{Edwards:1995,Wood:1997zq} and polarized M\o ller scattering~\cite{Anthony:2005pm}. The extracted value for $Q_W({\rm Cs})$ has shifted very recently from a 1$\sigma$ deviation to perfect agreement with the SM. This is due to the state of the art atomic structure calculation of Ref.~\cite{Porsev:2009pr} which also brought the atomic theory uncertainty below the measurement error. This is of significant importance for $Z'$ studies, since they easily affect and conversely are strongly constrained by precision weak charges. A previous 2~$\sigma$ deviation in $Q_W({\rm Cs})$ based on the same measurement~\cite{Wood:1997zq} but a different evaluation of the atomic physics~\cite{Ginges:2003qt} even indicated the presence of a $Z'$~\cite{Erler:1999nx}. Related to nuclear weak charges are the two linear combinations of four-Fermi couplings $C_{1u}$ and $C_{1d}$ (with $\tan\gamma \approx 0.445$~\cite{Young:2008}) which are the result of a global analysis of parity violating 
electron scattering experiments on nuclear fixed targets~\cite{Young:2007zs}. 

\begin{table}[t!]
\centering
\begin{tabular}{|l|c|c|c|r|}
\hline Quantity & Group(s) & Value & Standard Model & pull \\ 
\hline
$M_Z$ \hspace{10pt}      [GeV]             &     LEP 1     &$ 91.1876 \pm 0.0021 $&$ 91.1874 \pm 0.0021  $&$ 0.1 $  \\
$\Gamma_Z$ \hspace{14pt} [GeV]      &     LEP  1    &$  2.4952 \pm 0.0023 $&$  2.4954 \pm 0.0010  $&$-0.1 $  \\
$\sigma_{\rm had}$ \hspace{11pt}[nb] &     LEP   1   &$ 41.541  \pm 0.058  $&$ 41.483 \pm 0.008  $&$ 1.6 $  \\
$R_e$                                                       &     LEP  1    &$ 20.804  \pm 0.050  $&$ 20.736 \pm 0.010  $&$ 1.4 $  \\
$R_\mu$                                                  &     LEP  1    &$ 20.785  \pm 0.033  $&$ 20.736 \pm 0.010  $&$ 1.5 $  \\
$R_\tau$                                                  &     LEP  1    &$ 20.764  \pm 0.045  $&$ 20.782 \pm 0.010  $&$-0.4 $  \\
$A_{FB} (e)$                                           &     LEP  1    &$  0.0145 \pm 0.0025 $&$ 0.0163  \pm 0.0002  $&$-0.7 $  \\
$A_{FB} (\mu)$                                       &     LEP  1    &$  0.0169 \pm 0.0013 $&$                ~    $&$ 0.5 $  \\
$A_{FB} (\tau)$                                       &     LEP   1   &$  0.0188 \pm 0.0017 $&$                ~    $&$ 1.5 $  \\
\hline
$R_b$                                                      &  LEP 1 + SLD   &$  0.21629\pm 0.00066  $&$  0.21578 \pm 0.00005  $&$ 0.8   $ \\
$R_c$                                                      &  LEP 1 + SLD   &$  0.1721 \pm 0.0030 $&$  0.17224 \pm 0.00003  $&$ 0.0  $ \\
$R_{s,d}/R_{(d+u+s)}$                          &    OPAL      &$  0.371  \pm 0.022  $&$  0.3592  $&$ 0.5   $ \\
$A_{FB} (b)$                                           &     LEP  1     &$  0.0992 \pm 0.0016 $&$  0.1033 \pm 0.0007  $&$-2.5   $ \\
$A_{FB} (c)$                                           &     LEP  1    &$  0.0707 \pm 0.0035 $&$  0.0738 \pm 0.0006  $&$-0.9   $ \\
$A_{FB} (s)$                                           & DELPHI + OPAL &$  0.098  \pm 0.011  $&$  0.1034 \pm 0.0001  $&$-0.5   $ \\
$A_b$                                                      &     SLD      &$  0.923  \pm 0.020  $&$  0.9347 \pm 0.0001  $&$-0.6   $ \\
$A_c$                                                      &     SLD      &$  0.670  \pm 0.027  $&$  0.6679 \pm 0.0004   $&$ 0.1   $ \\
$A_s$                                                      &     SLD      &$  0.895  \pm 0.091  $&$  0.9357 \pm 0.0001  $&$-0.4   $ \\
$Q_{FB}$                                                &     LEP 1     &$ 0.0403 \pm 0.0026  $&   $0.0423 \pm    0.0003$   &$-0.8  $  \\
\hline
$A_{LR}$ (hadrons)                               &     SLD      &$  0.1514 \pm 0.0022 $&$  0.1473 \pm 0.0010  $&$ 1.9  $  \\
$A_{LR}$ (leptons)                                &     SLD      &$  0.1544 \pm 0.0060 $&                                          &$ 1.2  $  \\
$A_\mu$                                                  &     SLD      &$  0.142  \pm 0.015  $&                                            &$ -0.4 $  \\
$A_\tau$                                                  &     SLD      &$  0.136  \pm 0.015  $&                                            &$ -0.8 $  \\
$A_e (Q_{LR})$                                     &     SLD      &$  0.162  \pm 0.043  $&                                             &$ 0.3  $  \\
$A_\tau ({\cal P}_\tau)$                         &     LEP  1    &$  0.1439 \pm 0.0043 $&                                          &$-0.8  $  \\
$A_e ({\cal P}_\tau)$                             &     LEP  1   &$  0.1498 \pm 0.0049 $&                                          &$ 0.5  $  \\
$\sin^2\theta_W^{\rm eff}(e)$              & Tevatron  &$ 0.2316 \pm  0.0018$ &$ 0.2315 \pm  0.0001$ &  $0.1  $  \\
\hline
\end{tabular}
\caption[]{$Z$-pole precision observables from LEP~1, the SLC, and the Tevatron. The SM errors are parametric as in Table~\ref{Non-Zpole}.}
\label{Zpole}
\end{table}

Finally, the constraints in the last two lines from first row CKM matrix unitarity~\cite{Amsler:2008zzb,Hardy:2008gy,Bossi:2008aa} and from the anomalous magnetic moment of the muon~\cite{Bennett:2004pv} are affected by $Z'$ bosons at the one-loop level. These loop diagrams are finite and give rise to rather small but not necessarily negligible effects. For example, a $W-Z'$ box contribution could violate quark-lepton universality in the charged-current sector (and therefore apparently violate CKM unitarity) and is logarithmically enhanced for large $M_{Z'}$~\cite{Marciano:1987ja}. We included the analogous effect in the parameter $\Delta r$~\cite{Sirlin:1980nh}
describing the relation between the Fermi constant and $M_W$.

The size of the mixing angle $\theta_{ZZ'}$ is strongly constrained by the very high precision $Z$-pole experiments~\cite{EWWG:2005ema} at LEP and SLC shown in Table~\ref{Zpole}. The first set of measurements is from the $Z$ line shape, from the (inverse) leptonic branching ratios normalized to the total hadronic $Z$ decay width and from leptonic forward-backward asymmetries, $A_{FB}(\ell)$. The second set represents similarly defined quantities of the quark sector. While all $A_{FB}$ are practically sensitive only to the effective weak mixing angle defined for the initial state, $\sin^2\theta_W^{\rm eff}(e)$, the quantities $A_q$ are functions of the effective weak mixing angle of the respective quark flavor, $q$ ($Q_{FB}$ is a similar observable for light quarks). The third set is a variety of cross section asymmetries sensitive to $\sin^2\theta_W^{\rm eff}(e)$, $\sin^2\theta_W^{\rm eff}(\mu)$, or $\sin^2\theta_W^{\rm eff}(\tau)$. For details, see references~\cite{EWWG:2005ema,Erler:1998df}. The most recent result is the determination of $\sin^2\theta_W^{\rm eff}(e)$ by the CDF~\cite{Acosta:2004wq} and D\O~\cite{Abazov:2008xq} Collaborations and is obtained from the forward-backward asymmetry for $e^+ e^-$ final states. Many of the entries in Table~\ref{Zpole} are of much higher precision than typical low-energy observables. The $Z'$ amplitude, however, is almost entirely out of phase with and therefore negligible compared to the resonating $Z$ amplitude.  The $Z'$ enters here mainly through a modification of the couplings of the ordinary $Z$ to fermions, as well as through Eq.~(\ref{wzrelation}), and indirectly by affecting the extracted value of the QCD coupling.


\begin{table}[t]
\centering
\begin{tabular}{|c||r|r|r|r||r|c|c||c|} 
\hline
 $Z'$ & \multicolumn{4}{c||}{$M_{Z'}$ [GeV]} & \multicolumn{3}{c||}{$\sin\theta_{ZZ'}$}  & $\chi^2_{\rm min}$ \\ \hline
& EW (this work) & CDF & D\O\ & LEP~2 & $\sin\theta_{ZZ'}$ & $\sin\theta_{ZZ'}^{\rm min}$ & $\sin\theta_{ZZ'}^{\rm max}$ & \\ 
\hline\hline 
$Z_\chi$         & 1,141\phantom{OOO} &    892 & 640 &    673 & $-0.0004$ & $-0.0016$ & 0.0006 & 47.3 \\ \hline
$Z_\psi$         &    147\phantom{OOO} &    878 & 650 &    481 & $-0.0005$ & $-0.0018$ & 0.0009 & 46.5 \\ \hline
$Z_\eta$         &    427\phantom{OOO} &    982 & 680 &    434 & $-0.0015$ & $-0.0047$ & 0.0021 & 47.7 \\ \hline
$Z_I$              & 1,204\phantom{OOO} &    789 & 575 &            & $ 0.0003$ & $-0.0005$ & 0.0012 & 47.4 \\ \hline
$Z_S$            & 1,257\phantom{OOO} &    821 &         &            & $-0.0003$ & $-0.0013 $& 0.0005 & 47.3 \\ \hline
$Z_N$            &    623\phantom{OOO} &    861 &         &            & $-0.0004$ & $-0.0015$ & 0.0007 & 47.4 \\ \hline
$Z_R$            &    442\phantom{OOO} &            &         &            & $-0.0003$ & $-0.0015$ & 0.0009 & 46.1 \\ \hline
$Z_{LR}$       &    998\phantom{OOO} &    630 &         &    804 & $-0.0004$ & $-0.0013$ & 0.0006 & 47.3 \\ \hline
$Z_{\not{L}}$ &   (803)\phantom{IOO} & (740) \hspace{-8pt} & & & $-0.0015$ & $-0.0094$ & 0.0081 & 47.7 \\ \hline
$Z_{SM}$      & 1,403\phantom{OOO} & 1,030 & 780 & 1,787 & $-0.0008$ & $-0.0026$ & 0.0006 & 47.2 \\ \hline
$Z_{string}$  & 1,362\phantom{OOO} &            &         &            & $ 0.0002$ & $-0.0005$ & 0.0009 & 47.7 \\ \hline\hline
SM                  & \multicolumn{4}{c||}{$\infty$}                               & \multicolumn{3}{c||}{0}                    & 48.5 \\ \hline
\end{tabular}
\caption{95\% C.L. lower mass limits on extra $Z'$ bosons for various models from EW precision data and constraints on $\sin\theta_{ZZ'}$ assuming $\rho_0 = 1$ (fixed). For comparison, we show (where applicable) in the third, fourth and fifth column the limits obtained by CDF, D\O\  and LEP~2. In the following columns we give, respectively, the central value and the 95\% C.L. lower and upper limits for $\sin \theta_{ZZ'}$. Also indicated is the $\chi^2$ minimum for each model. The last row is included for comparison with the standard case of only one $Z$ boson.}
\label{Limits}
\end{table}

\section{Results and Discussion}
\label{results}
In Table~\ref{Limits} we present our limits on the $Z'$ parameters for the models introduced in Section~\ref{intro}. In this Table we specify our results for the case $\rho_0 = 1$ fixed but make no further assumptions regarding the Higgs sector except that the Higgs boson mass, $M_H$, is restricted to  $114.4~{\rm GeV} \leq M_H \leq 1$~TeV, where the upper end is from requiring perturbativity. The lower limit is the SM bound from LEP~2~\cite{Barate:2003sz} although we recall that this does not necessarily apply in the presence of new physics. Also shown in Table~\ref{Limits} are the current limits on various $Z'$ boson masses from the Tevatron and LEP~2. The CDF limits~\cite{Aaltonen:2008ah} are from a search for a dimuon invariant mass peak. Notice that the $Z_I$ and $Z_{\not L}$ bosons face the weakest limits as is expected from their hadrophobic and leptophobic characters, respectively (no limit on the $Z_{LR}$ is available from Run~II at the Tevatron; the entry shown in the Table is the CDF Run~I result~\cite{Abe:1997fd} from the combined dimuon and dielectron channels). Not shown are the dielectron channel search limits from CDF Run II~\cite{Aaltonen:2008vx} which are similar but slightly lower.  There is a significant excess at a dielectron invariant mass of 240~GeV, but this is not confirmed in the $\mu^+ \mu^-$ channel. The results from D\O~\cite{Hooper:2005gf} are based on the dielectron final state. The mass limits at the Tevatron assume that no decay channels into exotic fermions or superpartners are open to the $Z'$; otherwise the limits would be moderately weaker. LEP~2 constrains virtual $Z'$ bosons by their effects on cross sections and angular distributions of dileptons, hadrons, $b\bar{b}$ and $c\bar{c}$ final states~\cite{Alcaraz:2006mx}. The Table shows that the mass limits from the EW precision data are generally competitive with and in many cases stronger than those from colliders.  We stress that these classes of limits are highly complementary. The result for the leptophobic $Z_{\not{L}}$ (in parentheses) in the EW column is for the special Higgs sector with $\tau = 1/2$, {\em i.e.}, for the lower end of the restricted range in Table~\ref{cranges}.  For the upper end ($\tau = 1$) we find a limit of 1.32~TeV. The CDF number~\cite{Aaltonen:2008dn} refers to the $Z_{SM}$ limit from the dijet channel and should give a rough estimate of the sensitivity to our specific $Z_{\not{L}}$.

In the most general situation $\rho_0$ is allowed to differ from~1 and is treated as a free fit parameter. We give the results of this case for the $Z_\psi$ and the $Z_R$ models in Table~\ref{RhoFree}. The comparison with Table~\ref{Limits} shows that the presence of the extra fit parameter has little impact on the extracted $Z'$ constraints. 

Note, that all weak charges and the $C_{1q}$ are proportional to some vector coupling, $v$, and hence blind to the $Z_\psi$ which has only axial-vector couplings, $a$, to ordinary fermions.  This is why the EW data give very weak constraints on its mass. The loop effects on the last two observables in Table~\ref{Non-Zpole} gain therefore relative importance. In fact, $Z'$ effects on $g_\mu - 2$ are proportional to $v_\mu^2 - 5\, a_\mu^2$ so that there is an additional enhancement (of an otherwise very small effect).

\begin{table}[t]
\centering 
\begin{tabular}{|c||c||c|c|c||c|c|c||c|} \hline
$Z'$ & $M_{Z'}$ [GeV] & $\sin \theta_{ZZ'}$ & $\sin\theta_{ZZ'}^{\rm min}$ & $\sin\theta_{ZZ'}^{\rm max}$ & $\rho_0$ & $\rho_0^{\rm min}$ & $\rho_0^{\rm max}$ & $\chi^2_{\rm min}$ \\ \hline
\hline
$Z_\psi$ & 147 & $-0.0004$ & $-0.0018$ & 0.0010 & 1.0002 & 0.9996 & 1.0035 & 46.1 \\ \hline
$Z_R$    & 439 & $-0.0003$ & $-0.0015$ & 0.0012 & 1.0003 & 0.9996 & 1.0035 & 45.3 \\ \hline\hline
SM & $\infty$ & \multicolumn{3}{c||}{0} & 1.0003 & 0.9996 & 1.0035 & 47.9 \\ \hline
\end{tabular}
\caption{95\% C.L. limits on $M_Z'$, $\sin \theta_{ZZ'}$ and $\rho_0$ when the latter is allowed to float freely.}
\label{RhoFree}
\end{table}

\begin{figure}[h!]
\vspace{-96pt}
\centering 
\begin{tabular}{cc}
\includegraphics[scale=0.38]{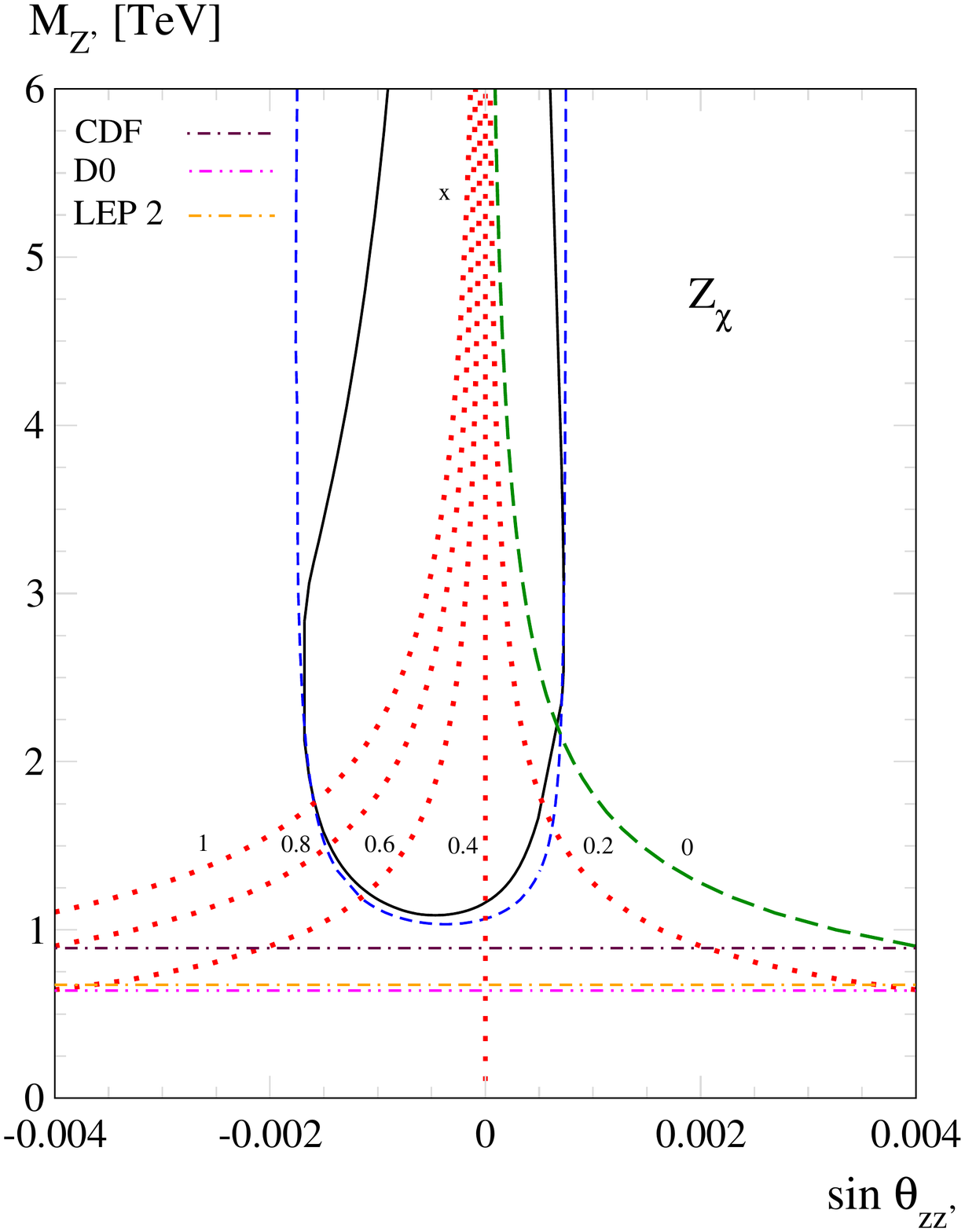} & \includegraphics[scale=0.38]{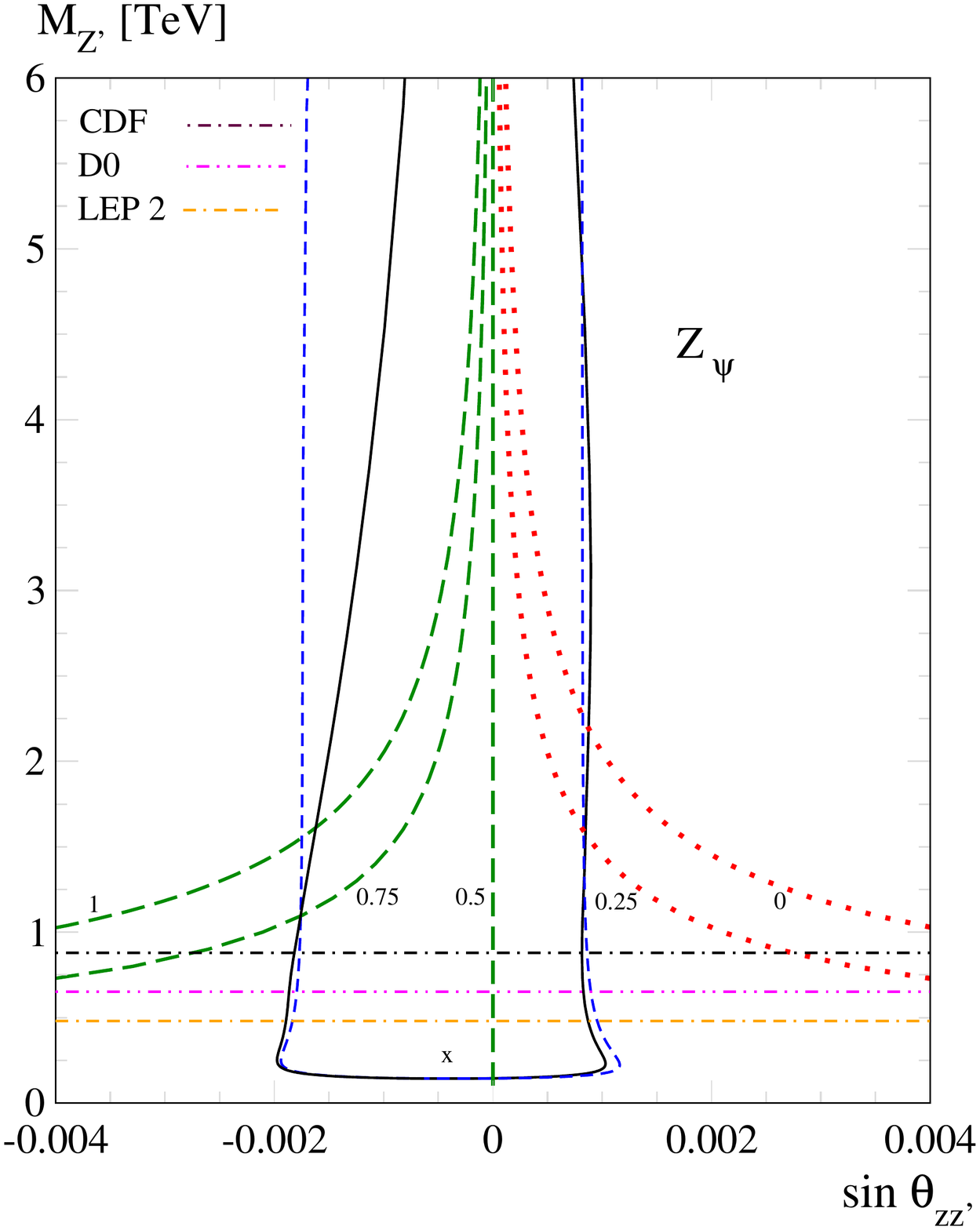} \vspace{-12pt} \\
\includegraphics[scale=0.38]{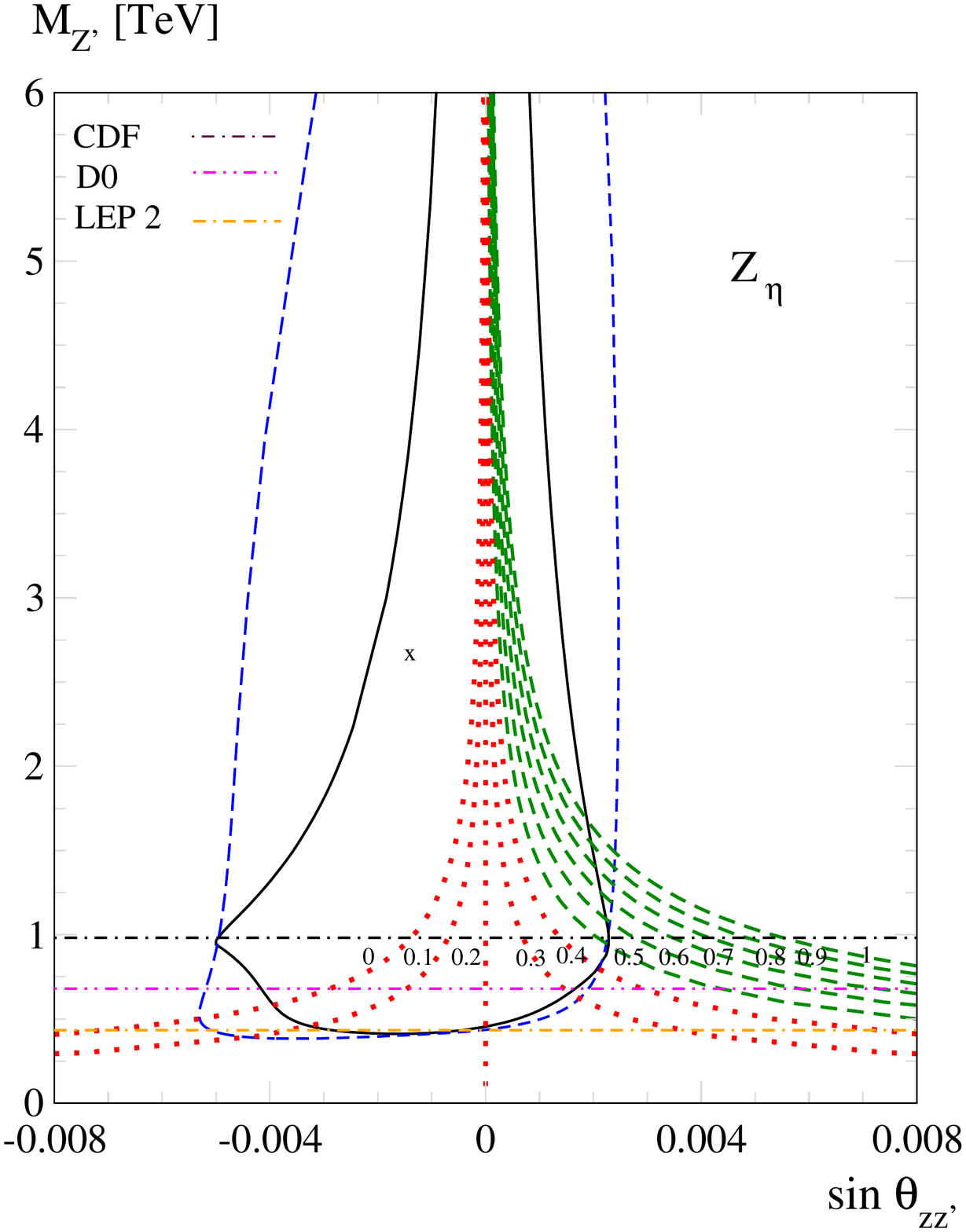} & \includegraphics[scale=0.38]{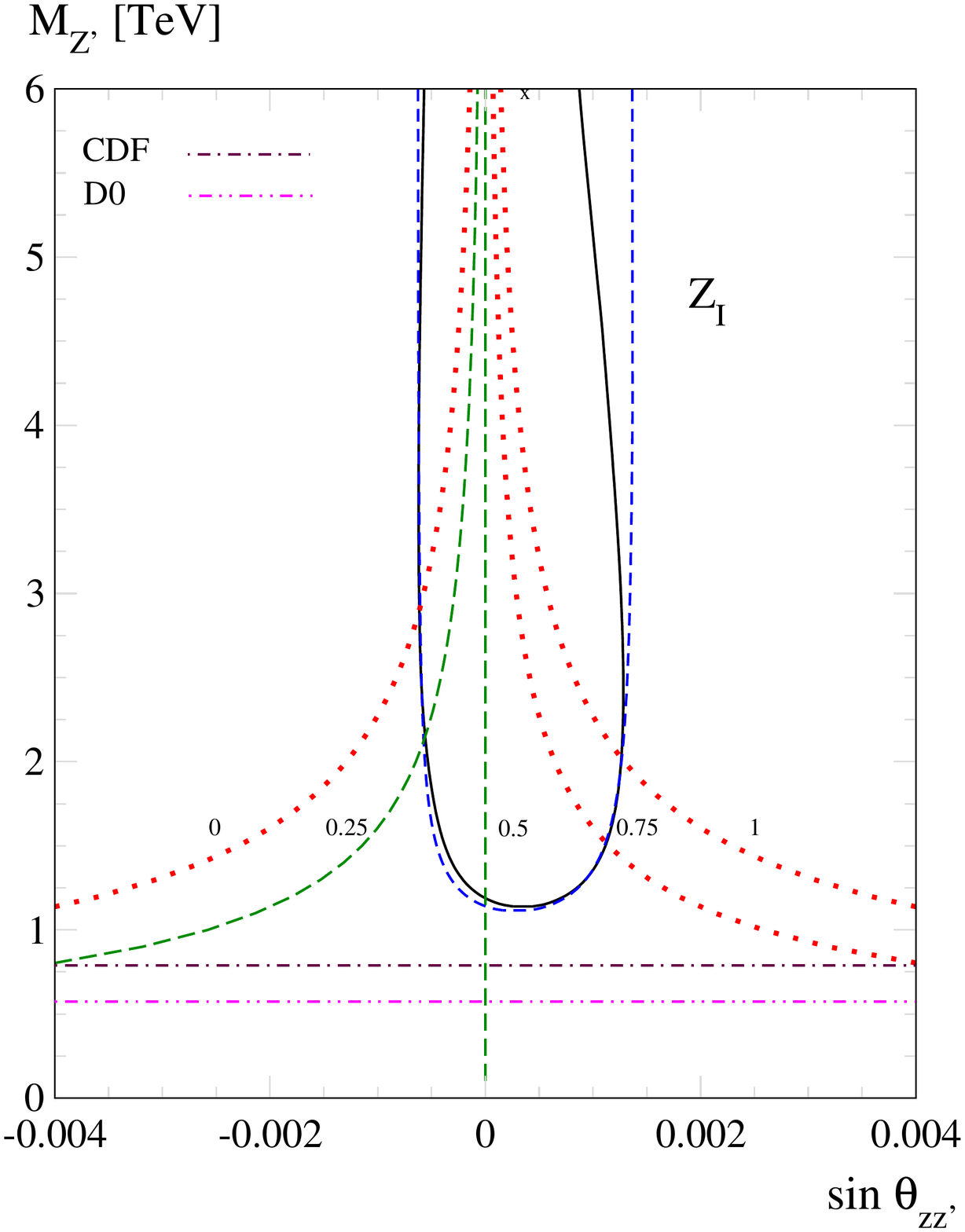} \\
\end{tabular}
\caption{95\% C.L. contours in $M_{Z'}$ vs. $\sin \theta_{ZZ'}$ for various models. See the text for details.}
\label{Contours}
\end{figure}

\begin{figure}[h!]
\vspace{-96pt} 
\centering
\begin{tabular}{cc}
\includegraphics[scale=0.38]{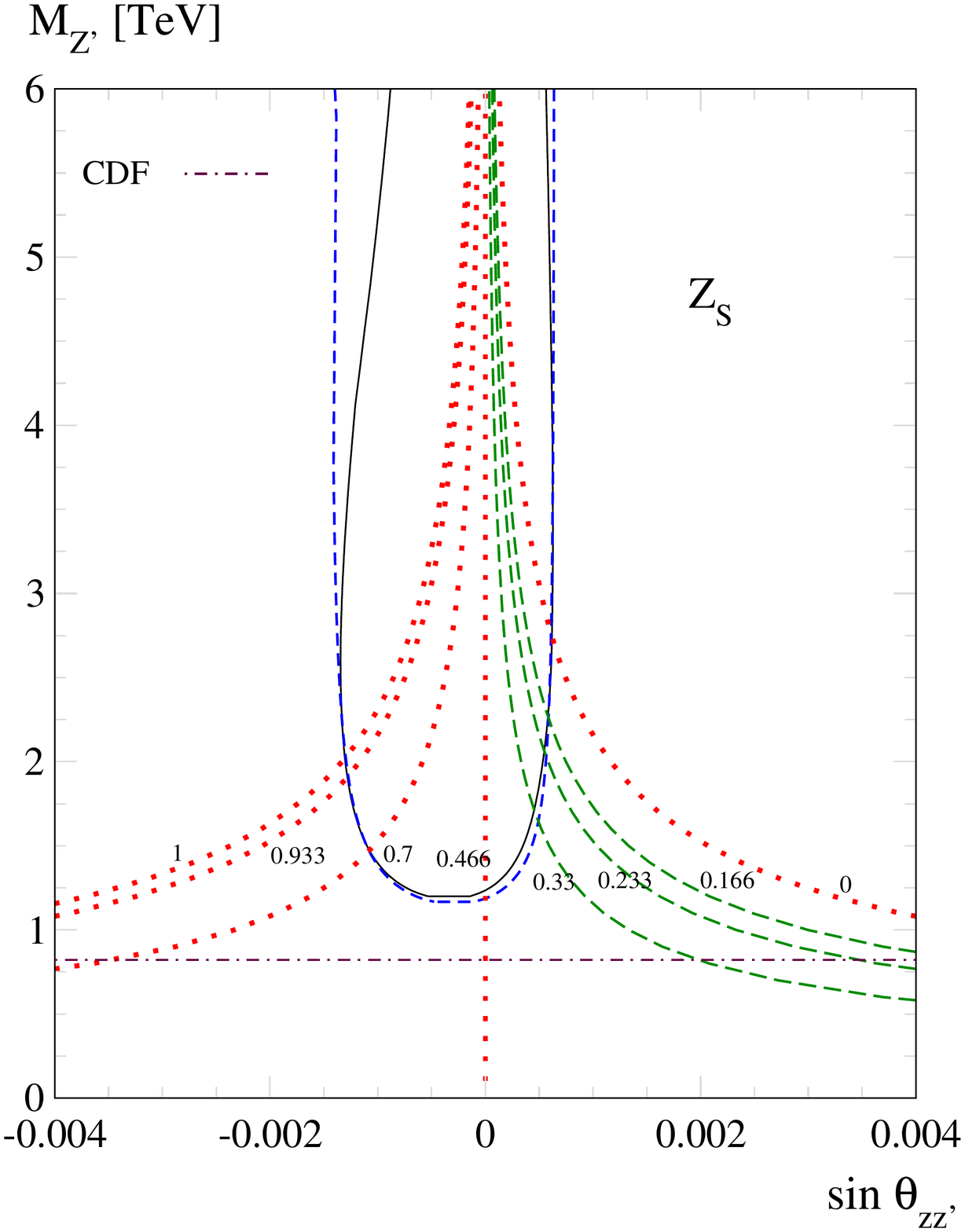}    & \includegraphics[scale=0.38]{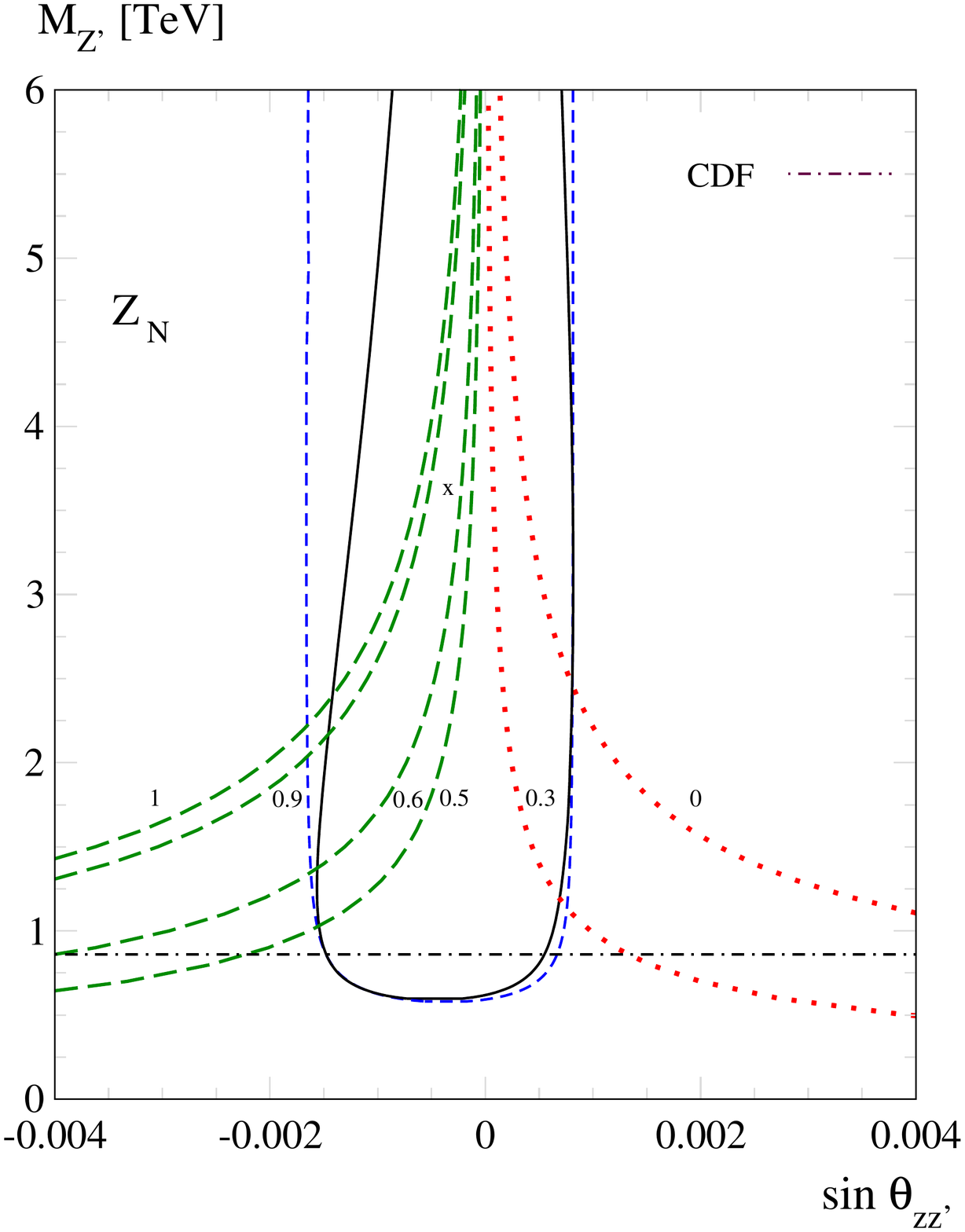} \vspace{-12pt} \\
\includegraphics[scale=0.38]{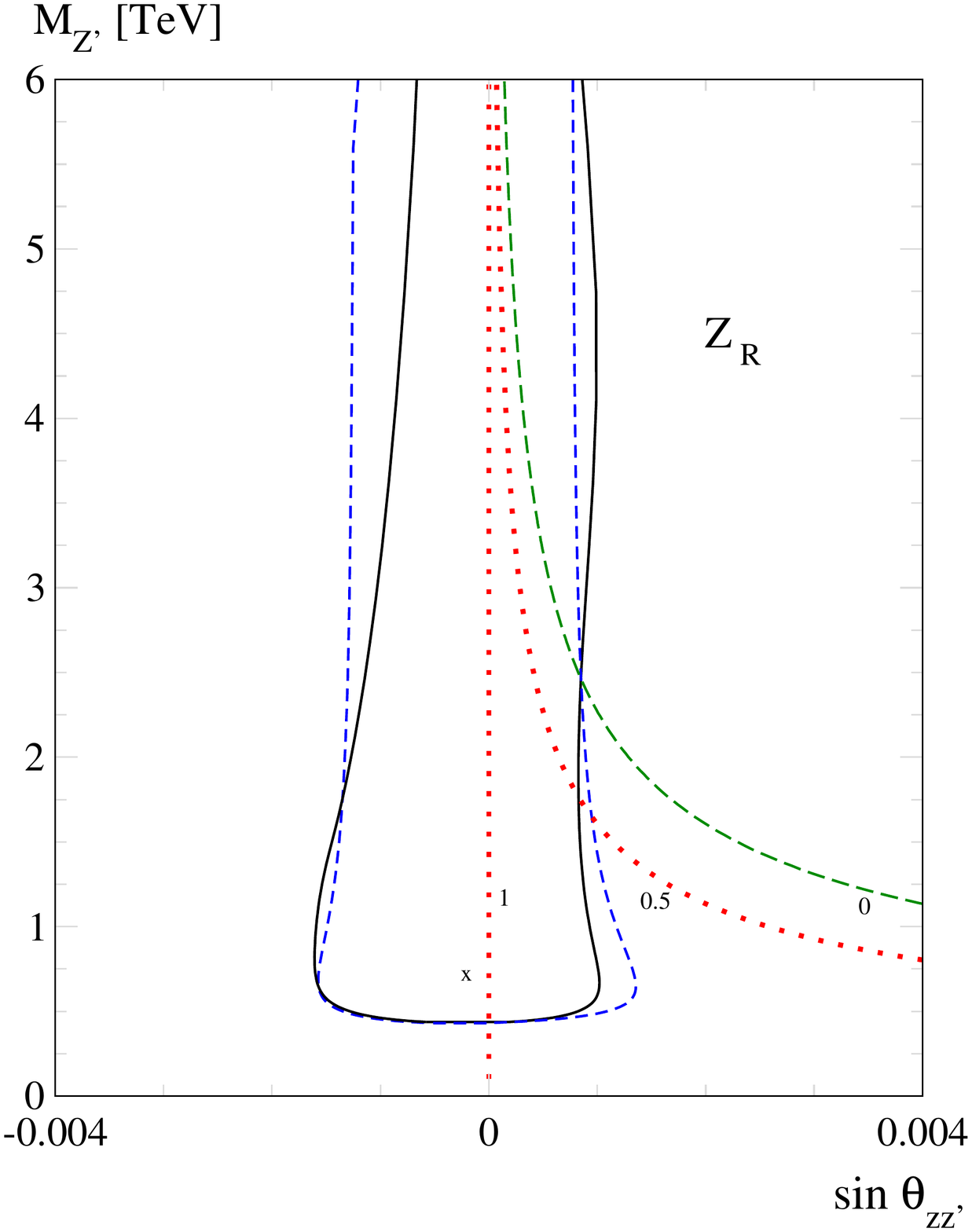} & \includegraphics[scale=0.38]{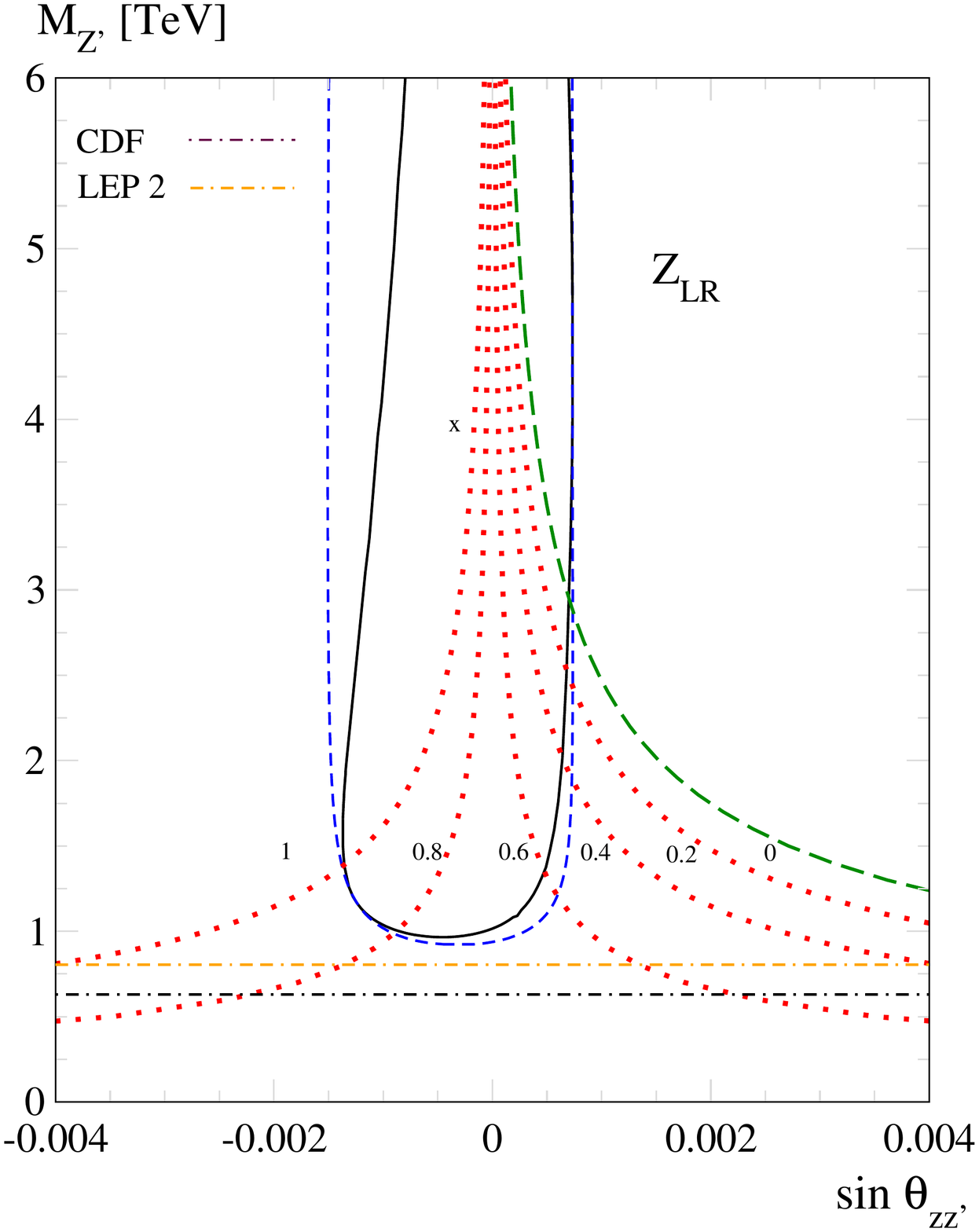} \\
\end{tabular}
\caption{95\% C.L. contours in $M_{Z'}$ vs. $\sin \theta_{ZZ'}$ for various models. See the text for details.}
\label{Contours2}
\end{figure}

\begin{figure}[h!]
\vspace{-96pt} 
\centering
\begin{tabular}{cc}
\includegraphics[scale=0.38]{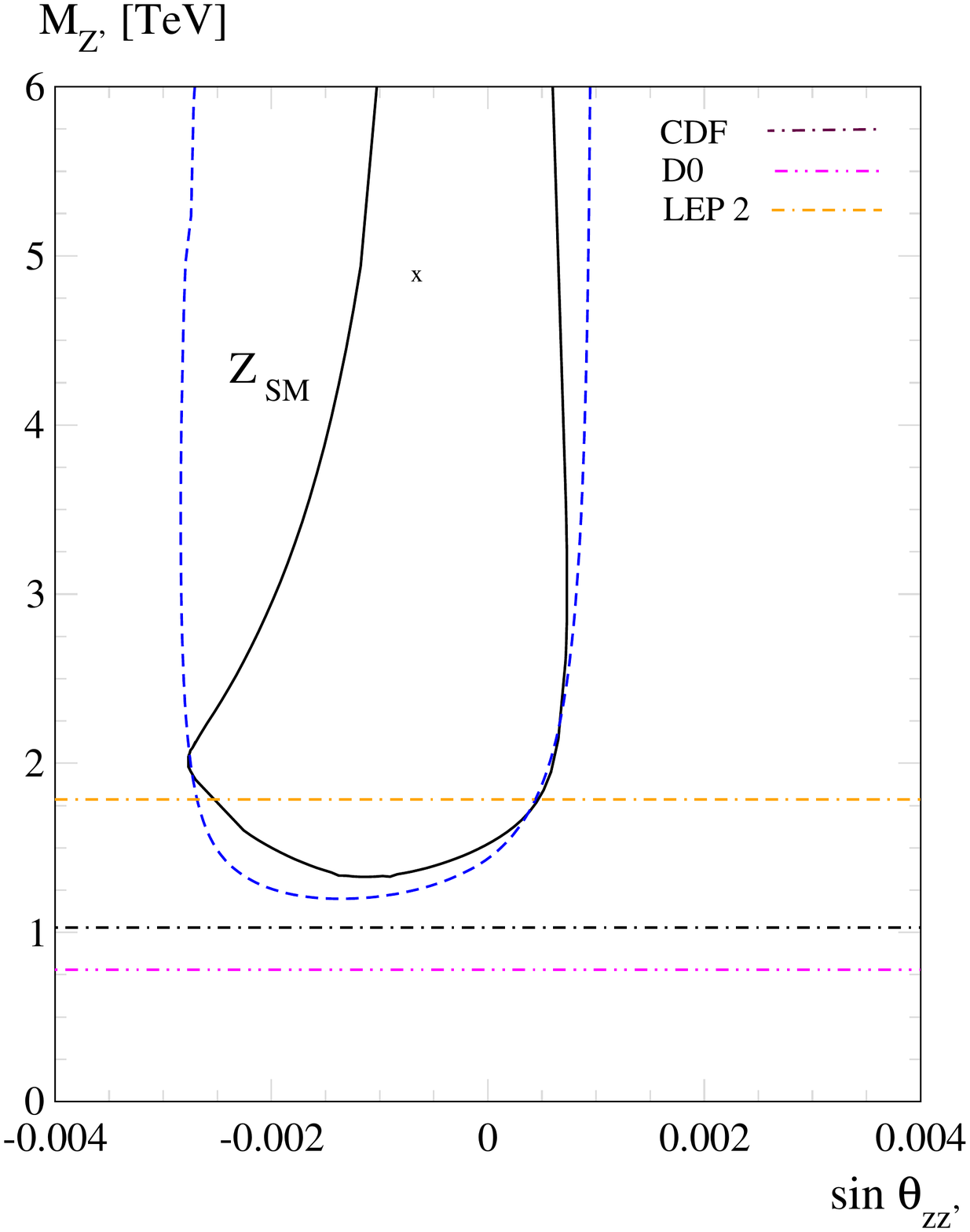} & \includegraphics[scale=0.38]{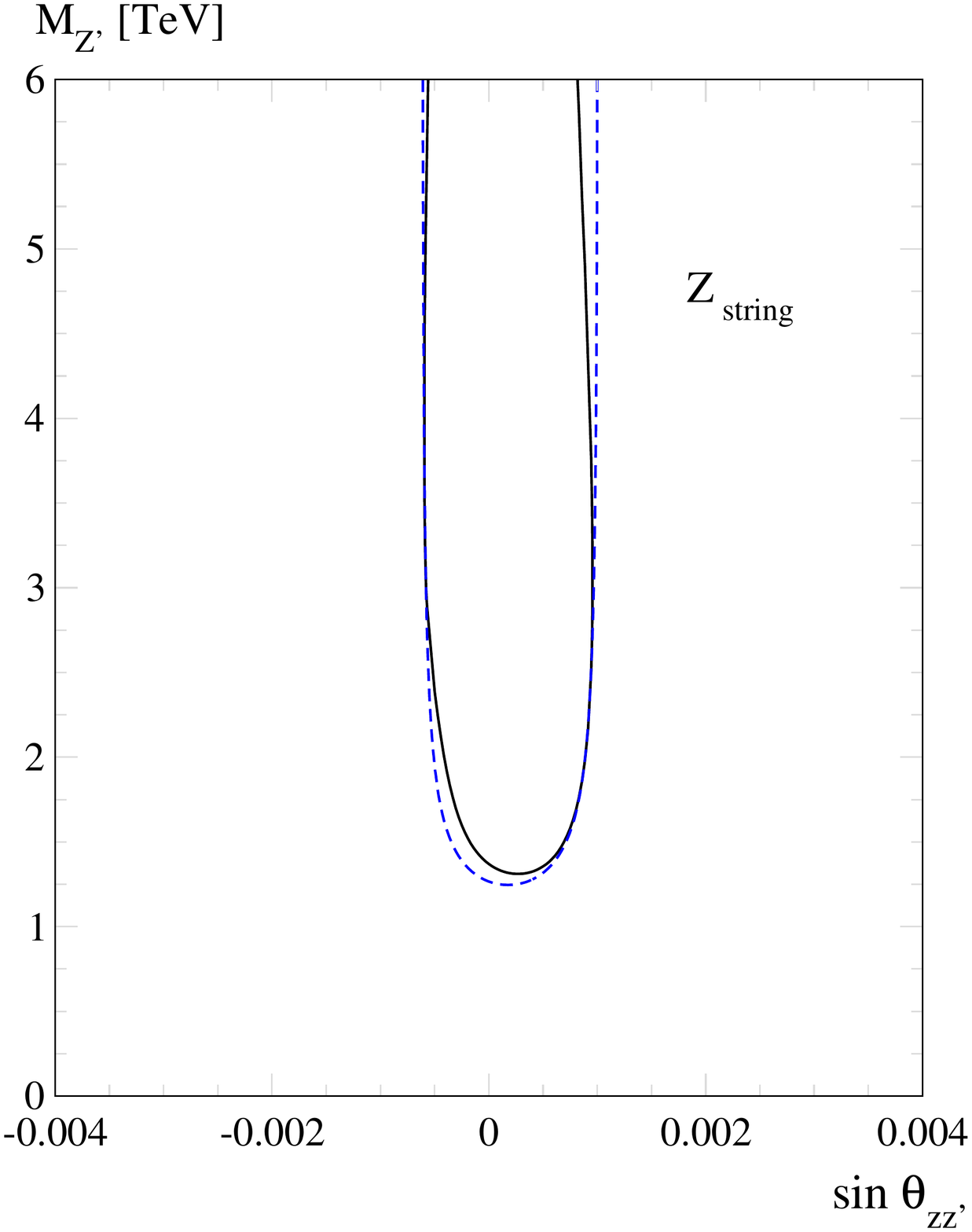} \\
\end{tabular}
\caption{95\% C.L. contours in $M_{Z'}$ vs. $\sin \theta_{ZZ'}$ for the sequential $Z'$ boson and the $Z_{string}$ model. See the text for details.}
\label{Contours3}
\end{figure}

Figures~\ref{Contours}, \ref{Contours2} and~\ref{Contours3} show 90\% C.L. exclusion contours for all models except for the $Z_{\not L}$
(since its mass is in general unbounded). The solid (black) lines specify use of the constraint $\rho_0 = 1$ while the dashed (blue) lines are for $\rho_0$ free. We also show the extra constraints for the specific Higgs sectors described in Section~\ref{exotics}. These are represented by the dotted (red) lines unless they belong to the restricted range in Table~\ref{cranges} in which case they are long-dashed (green). The numbers in the plots refer to the values of $\tau$ or $\omega$, whichever carries the larger coefficient in Table~\ref{cranges}. The best fit locations (for $\rho_0 = 1$) are indicated by an {\tt "x"}. The lower limits from CDF (dot-dashed and black), D\O\ (double-dot-dashed and magenta) and LEP~2 (dot-double-dashed and orange) given in Table~\ref{Limits} are also shown. 

\begin{table}[b]
\centering 
\begin{tabular}{|c||c|c|c|c|c|c|} \hline
$Z'$ & $Z_\chi$ & $Z_\psi$ & $Z_\eta$ & $Z_I$ & $Z_S$ & $Z_N$ \\ \hline
& & & & & &\vspace{-6pt} \\
$M_H$ [GeV] & $171^{+493}_{-\phantom{4}89}$ & $97^{+31}_{-25}$                           & $423^{+577}_{-350}$                    & 
$141^{+304}_{-\phantom{3}61}$ & $149^{+353}_{-\phantom{3}68}$ & $117^{+222}_{-\phantom{2}40}$ \\ \vspace{-6pt}
& & & & & &\\  \hline
$\chi^2_{\rm min}$ & 47.3 & 46.1 & 47.7 & 47.4 & 47.3 & 47.4 \\ \hline\hline
$Z'$ & $Z_R$ & $Z_{LR}$ & $Z_{\not L}$ & $Z_{SM}$ & $Z_{string}$ & SM \\ \hline
& & & & & &\vspace{-6pt} \\
$M_H$ [GeV]  & $84^{+31}_{-24}$                           & $110^{+174}_{-\phantom{1}35}$ & $126^{+276}_{-\phantom{2}52}$ &
$331^{+669}_{-246}$                    & $134^{+299}_{-\phantom{2}58}$ &  $96^{+29}_{-25}$                          \\ \vspace{-6pt}
& & & & & & \\  \hline
$\chi^2_{\rm min}$ & 45.1 & 47.3 & 47.7 & 47.2 & 47.7 & 48.0 \\  \hline
\end{tabular}
\caption{$1\sigma$ ranges of $M_H$ allowed by each model and the best fit $\chi^2$ values.}
\label{MH}
\end{table}

In all Figures and in Tables~\ref{Limits} and~\ref{RhoFree} we used the $M_H$ window mentioned above. However, the SM best fit value, $M_H =  96^{+29}_{-24}$~GeV, is below this range. It is interesting to note~\cite{Erler:1999nx,Chanowitz:2008ix} that the presence of a $Z'$ often moves the central value up to the allowed region. Table~\ref{MH} shows the best fit values and $1\sigma$ errors for $M_H$ when the LEP~2 bound is removed. 

Some $Z'$ models have a fairly low minimum $\chi^2$, especially the $Z_\psi$ and the $Z_R$. Table~\ref{MH} shows the $\chi^2$ minimum of the $Z_R$ model about 3~units below the SM value, technically implying an {\em upper\/} bound on the $Z_R$ mass of about 29 TeV at the 90\% C.L. This is actually the reason why we included the $Z_R$ in this paper in the first place.  Of course, at present there is little significance to this observation since we have two additional fit parameters ($M_Z'$ and $\sin \theta_{ZZ'}$) and various parameters for the charges (like the angles $\alpha$ and $\beta$) to adjust. Nevertheless, this is somewhat surprising given that the SM fit is quite good with $\chi^2_{\rm min} = 48.0$ for 45 effective degrees of freedom.  It may be useful to note that the improvement in $\chi^2$ arises mainly through $\sigma_{\rm had}$, $Q_W(e)$, and the $e^-$-DIS observables, where the latter two are of special interest in view of proposed and approved experiments to be performed at JLab. 

\section*{Acknowledgements} 
We thank the Institute for Nuclear Theory at the University of Washington for its hospitality and the Department of Energy for partial support during the early stages of this work.  Work at IF-UNAM is supported by CONACyT project 82291--F. The work of P.L.~is supported by the IBM Einstein Fellowship and by NSF grant PHY--0503584.

\end{document}